\documentclass[aps,prd,preprint,superscriptaddress,tightenlines,nofootinbib]{revtex4}

\usepackage{graphicx}
\usepackage{dcolumn}
\usepackage{bm}

\newcommand{\pp}{\psi(2S)}   
\newcommand{\pdp}{\psi(3770)}
\newcommand{\DDbar}{D \bar D}

\newcommand{\FourPi}   {$2(\pi^+ \pi^-)$}

\newcommand{\FivePi}   {$2(\pi^+ \pi^-)\pi^0$}
\newcommand{\EtaPiPi}  {$\eta \pi^+ \pi^-$}
\newcommand{\OmegaPiPi}{$\omega \pi^+ \pi^-$}

\newcommand{\EtaTriPi} {$\eta 3\pi$}
\newcommand{\EtaTriPiA}{$\eta 3\pi$ $^a$}
\newcommand{\EtaTriPiB}{$\eta 3\pi$ $^b$}
\newcommand{\EtapTriPi}{$\eta^\prime 3\pi$}
\newcommand{\KKPiPi}   {$K^+K^-\pi^+\pi^-$}

\newcommand{\PhiPiPi}  {$\phi \pi^+ \pi^-$}
\newcommand{\PhiFzero} {$\phi f_0$}

\newcommand{\KKTriPi}  {$K^+K^-\pi^+\pi^-\pi^0$}
\newcommand{\EtaKK}    {$\eta K^+ K^-$}
\newcommand{\OmegaKK}  {$\omega K^+ K^-$}
\newcommand{\KKKK}     {$2(K^+K^-)$}
\newcommand{\PhiKK}    {$\phi K^+ K^-$}
\newcommand{\KKKKPi}   {$2(K^+K^-)\pi^0$}
\newcommand{\PPPiPi}   {$p \bar{p} \pi^+ \pi^-$}

\newcommand{\PPTriPi}  {$p \bar{p} \pi^+ \pi^-\pi^0$}
\newcommand{\EtaPP}    {$\eta p \bar{p}$}
\newcommand{\OmegaPP}  {$\omega p \bar{p}$}
\newcommand{\PPKK}     {$p \bar{p} K^+ K^-$}
\newcommand{\PhiPP}    {$\phi p \bar{p}$}
\newcommand{\LamLam}   {$\Lambda \bar\Lambda$}
\newcommand{\LLPiPi}   {$\Lambda\bar\Lambda\pi^+\pi^-$}
\newcommand{\LamPK}    {$\Lambda \bar{p} K^+$}
\newcommand{\LamPKPiPi}{$\Lambda\bar{p}K^+\pi^+\pi^-$}
\def\Journal#1&#2&#3(#4){#1{\bf #2}, #3 (#4)}
\def\NIMA{Nucl. Instrum. Methods Phys. Res., Sect. A }

\def\PLB{Phys.  Lett.  B }
\def\PRL{Phys.  Rev.  Lett.  }

\begin{document}
\preprint{CLNS 05/1921}       
\preprint{CLEO 05-13}         

\title{\bf Search for Exclusive Multi-body Non-$D\bar{D}$ Decays at the $\pdp$}

\author{G.~S.~Huang}
\author{D.~H.~Miller}
\author{V.~Pavlunin}
\author{B.~Sanghi}
\author{I.~P.~J.~Shipsey}
\affiliation{Purdue University, West Lafayette, Indiana 47907}
\author{G.~S.~Adams}
\author{M.~Cravey}
\author{J.~P.~Cummings}
\author{I.~Danko}
\author{J.~Napolitano}
\affiliation{Rensselaer Polytechnic Institute, Troy, New York 12180}
% ...
\collaboration{CLEO Collaboration} 
\noaffiliation

\date{January 27, 2006}

\begin{abstract}
Using data collected at the $\psi(3770)$ resonance with the CLEO-c
detector at the Cornell $e^+e^-$ storage ring, we present searches
for 25 charmless decay modes of the $\psi(3770)$, mostly
multibody final states. No evidence for charmless decays is found.
\end{abstract}

\maketitle

The $\pdp$ is the lowest-mass charmonium 
resonance above $\DDbar$ threshold \cite{PDG}. It may be the 
$1^3D_1$ state or a mixture of $1^3D_1$ and $2^3S_1$.
Charmless decays of the $\psi(3770)$ can shed light on $S-D$
mixing, missing $\psi(2S)$ decays such as $\psi(2S)\to\rho\pi$,
the discrepancy between the total and $D \bar D$ cross section at the 
$\psi(3770)$, and rescattering effects contributing to an enhanced
$b \to s$ penguin amplitude in $B$ meson decays \cite{rosner}.

The total cross section at the $\pdp$ was estimated from older measurements 
to be $(7.9\pm0.6)$ nb in \cite{rosner}, which is larger by about 20\%
than the measured $D \bar D$ cross section 
$(6.39\pm0.10^{+0.17}_{-0.08})$ nb \cite{ddxsec}.
While the significance of the discrepancy between the total cross section
and the $D \bar D$ cross section is not large, identifying
non-$D \bar D$ decays of the $\pdp$ will place the discrepancy on a solid
footing and shed light on the nature of the $\pdp$. The BES Collaboration
observed $\pdp\to\pi^+\pi^-J/\psi$ with a branching ratio of 
$(0.34\pm0.14\pm0.09)\%$ \cite{bespipijpsi}, while CLEO 
measured a branching ratio of $(0.189\pm0.022^{+0.007}_{-0.004})\%$
\cite{bkhhmk}.
This non-$D \bar D$ channel contributes approximately 100 keV
to the $\psi(3770)$ decay width, 
which motivates the search for other exclusive final states.

Charmless decays of the $\psi(3770)$ may provide an avenue to study
rescattering effect relevant to $B$ meson decays. For example, 
the $\eta^\prime$ exhibits enhanced production in charmless 
inclusive and exclusive $B$ meson decays which is not well
understood. If the $\psi(3770)$ decays to $D \bar D$ 
pairs which subsequently re-annihilate into non-charmed final states, 
a similar effect could contribute to enhanced $b \to s$ penguin
amplitudes in $B$ meson decays, including modes containing
an $\eta ^\prime$, i.e. 
$b \to c \bar{c} s \to q \bar{q} s$, where $q=(u,d,s)$.

Although the $\pdp$ is believed to be primarily the $1^3D_1$ 
state of the $c\bar{c}$ system, its large leptonic width 
indicates mixing with $S$-wave states, particularly the $\pp$. 
By enhancing the rate of non-$D \bar D$ decays at
the $\psi(3770)$, mixing between the $\psi(3770)$ and 
the $\psi(2S)$ provides an explanation for the anomalously
small $\psi(2S)$  branching fractions to some  hadronic 2-body 
final states such as $\rho \pi$ \cite{rosner}.

In this Letter, we report results of searches for a wide variety of 
non-$D \bar D$ modes,
including final states with and without strangeness and with and without
baryons. The modes $\eta(^\prime)3\pi$ are included since recent predictions
exist \cite{rosner}. Modes with baryons are included since any observation
would be unambiguously a non-$D \bar D$ decay because $D$ mesons are not
sufficiently massive to decay to baryon pairs.

The data sample used in this analysis is obtained at the $\pdp$ 
and the nearby continuum in $e^+e^-$ collisions produced by the 
Cornell Electron Storage Ring (CESR) and acquired with the CLEO-c
detector. The CLEO-c~\cite{YELLOWBOOK} detector is a modification 
of the CLEO III detector \cite{cleoiiidetector}, in which
the silicon-strip vertex detector was replaced by a six-layer
all-stereo drift chamber. The solid angle coverage for charged 
and neutral particles is 93\% of $4\pi$.
The charged particle tracking system, operating in a
1.0~T magnetic field directed along the beam axis, achieves
a momentum resolution of $\sim$0.6\% at
$p=1$~GeV. The calorimeter attains a photon
energy resolution of 2.2\% at $E_\gamma=1$~GeV and 5\% at 100~MeV.
Two particle identification systems, one based on energy loss ($dE/dx$) in
the drift chamber and the other a ring-imaging Cherenkov (RICH)
detector, are used together to identify $\pi^\pm$, $K^\pm$ and $p(\bar{p})$.
The combined $dE/dx$-RICH particle identification procedure has
efficiencies exceeding 90\% and misidentification rates below 5\%
for these three particle species.

The integrated luminosity ($\cal{L}$) of the datasets was measured
using $e^+ e^-$, $\gamma\gamma$, and $\mu^+ \mu^-$ final
states~\cite{LUMINS}. Event counts were normalized with a Monte
Carlo (MC) simulation based on the Babayaga~\cite{BBY} event
generator combined with GEANT-based~\cite{GEANT} detector modeling. 
The data consist of $\cal{L}$=55.8~pb$^{-1}$ on the peak of the 
$\pdp$ and 20.70~pb$^{-1}$ at the continuum ($\sqrt{s}$=3.67~GeV), 
which is used for background subtraction. 
The nominal scale factor used to normalize continuum yields 
to $\pdp$ data is $f_{\rm co}^{\rm nom}=2.55\pm0.26$, 
and is determined from the integrated luminosities of the data sets 
corrected for an assumed $1/s$ dependence of the cross section where the 
error is from the uncertainties in the relative luminosity and 
the $s$ dependence of the cross section. The scale factor 
differs by a factor of 5.2\% for each power of $1/s$.
The value of $f_{\rm co}$ used for each mode also corrects for the
small difference in efficiency between the $\psi(3770)$ and
continuum data.

The analysis strategy and selection criteria are the same as in the 
CLEO-c analysis of exclusive hadronic decays at the $\pp$ \cite{psipmulti}.
Standard requirements are used to select charged particles
reconstructed in the tracking system and photon candidates in the
CsI calorimeter. We require tracks of charged particles
to have momenta $p>100$~MeV
and to satisfy $|\cos\theta|<0.90$, where $\theta$ is the polar angle
with respect to the $e^+$ direction.
Each photon candidate satisfies  $E_\gamma>30$~MeV and is more than 8\,cm
away from the projections of charged tracks into the calorimeter.
Particle identification is used for all charged
particle candidates. Pions, kaons, and protons must be positively and
uniquely identified. That is: 
pion candidates must not satisfy kaon or proton selection criteria,
and kaon and proton candidates obey similar requirements.
Charged particles must not be identified as electrons using criteria
based on momentum, calorimeter energy deposition, and $dE/dx$.

The invariant mass of the decay products from the following particles
must lie within limits determined from MC studies:
$\pi^0~(120 \le M_{\gamma\gamma} \le 150 {\rm ~MeV})$, 
$\eta ~(500 \le M_{\gamma\gamma} \le 580 {\rm ~MeV}$, 
$530 \le M_{\pi^+\pi^-\pi^0} \le 565 {\rm ~MeV})$, 
$\omega ~(740 \le M_{\pi^+\pi^-\pi^0} \le 820 {\rm ~MeV}$ 
[$760 \le M_{\pi^+ \pi^- \pi^0} \le 800 {\rm ~MeV}$ 
for the $\omega p \bar p$ final state]), 
$\phi ~(1.00 \le M_{K^+K^-} \le 1.04 {\rm ~GeV})$, and 
$\Lambda ~(1.112 \le M_{p \pi^-} \le 1.120 {\rm ~GeV})$. 
For $\pi^0 \rightarrow \gamma \gamma$
and $\eta \rightarrow \gamma \gamma$ candidates in events with
more than two photons, the combination giving a mass closest to
the known  $\pi^0$ or $\eta$ mass is chosen, and a kinematically
constrained fit to the known parent mass is made \cite{kinefit}. 
To suppress electromagnetic energy deposits in the calorimeter
mimicking a $\pi^0$ or $\eta$, each electromagnetic shower profile
is required to be consistent with that of a photon.
For $\eta \rightarrow \pi^+ \pi^- \pi^0$ and 
$\omega \rightarrow \pi^+ \pi^- \pi^0$, 
the $\pi^0$ is selected as described above, and then
combined with all possible combinations of two oppositely charged
pions choosing the combination that is closest to the known $\eta
(\omega)$ mass. 
For $\Lambda \rightarrow p \pi^-$, a fit of the $p$ and $\pi^-$ 
trajectories to a common vertex separated from the $e^+e^-$ 
interaction ellipsoid is made. Contamination from $K^0_S$ 
decays is eliminated by particle identification and energy 
conservation requirements.

Reconstructed events must conserve momentum and energy. 
The hadrons comprising these events each have momentum $p_i$ 
and combined measured energy $E_{\rm vis}$. We require the measured
scaled energy $E_{\rm vis}$/$E_{\rm cm}$ be consistent with unity
within experimental resolution, which varies by final state.
We require $|\Sigma {\bf p_i}|/E_{\rm cm}< 0.02$. Together these
requirements suppress backgrounds with missing energy or incorrect
mass assignments. The experimental resolutions are smaller than
1\% in scaled energy and 2\% in scaled momentum.

For the final states with four charged tracks and a $\pi^0$,
an additional cut is applied to remove a background of radiative
events. When the highest energy photon in an event is combined 
with a low-energy photon candidate, it can imitate a $\pi^0$. 
We require $(E_{\rm 4 tracks}+E_\gamma)$/$E_{\rm cm} < 0.995$,
where $E_\gamma$ is the energy of the highest energy photon.
For the final states $2(\pi^+ \pi^-)$ and  $2(\pi^+ \pi^-)\pi^0$,
there is a background from $(\gamma) \pi^+ \pi^- J/\psi$ arising 
mostly from radiative returns to the $\psi(2S)$. This background 
is vetoed if the recoil mass against the two slowest oppositely 
charged tracks (assumed to be pions), $m_2^{\rm slow}$, satisfies 
$3.15 < m_2^{\rm slow} <3.22 {\rm ~GeV}$, and/or the invariant
mass of the two fastest oppositely charged tracks (assumed to be
muons unless the $dE/dx$ measurement is consistent with electrons),
$m_2^{\rm fast}$, satisfies $3.05 < m_2^{\rm fast} <3.15 {\rm ~GeV}$.
For the final states \FivePi, \KKPiPi\ and \KKTriPi, 
in order to remove $\DDbar$ background, we exclude events
in which the invariant mass of the following combinations
of particles is consistent with a $D^0/D^\pm$ meson: 
$\pi^+\pi^-\pi^0$, $K^\pm\pi^\mp$, $\pi^+\pi^-$, $K^+K^-$, 
$K^\pm\pi^\mp\pi^0$, $K^+K^-\pi^0$ or $\pi^\pm\pi^0$.

For every  final state, a signal selection range in 
$E_{\rm vis}$/$E_{\rm cm}$ is determined by Monte Carlo simulation, 
and a sideband selection range is defined to measure background. 
The signal range in $E_{\rm vis}$/$E_{\rm cm}$ varies between
$0.98-1.02$ and $0.99-1.01$ depending on the final state.
Final states with an intermediate $\eta$, $\omega$, or $\phi$
must satisfy a scaled energy signal selection range requirement
identical to the corresponding mode without the intermediate
particle, and the event yield is determined from signal and 
sideband selection ranges of the particle mass.  For example, 
the scaled energy signal selection range is
the same for $\phi K^+K^-$ and $K^+K^-K^+K^-$.
Most modes studied in this Letter have resonant
sub-modes, however, only narrow resonances are included
in this analysis.

\begin{figure}[htbp]
  \centering
  \includegraphics[height=0.35\textheight]
  {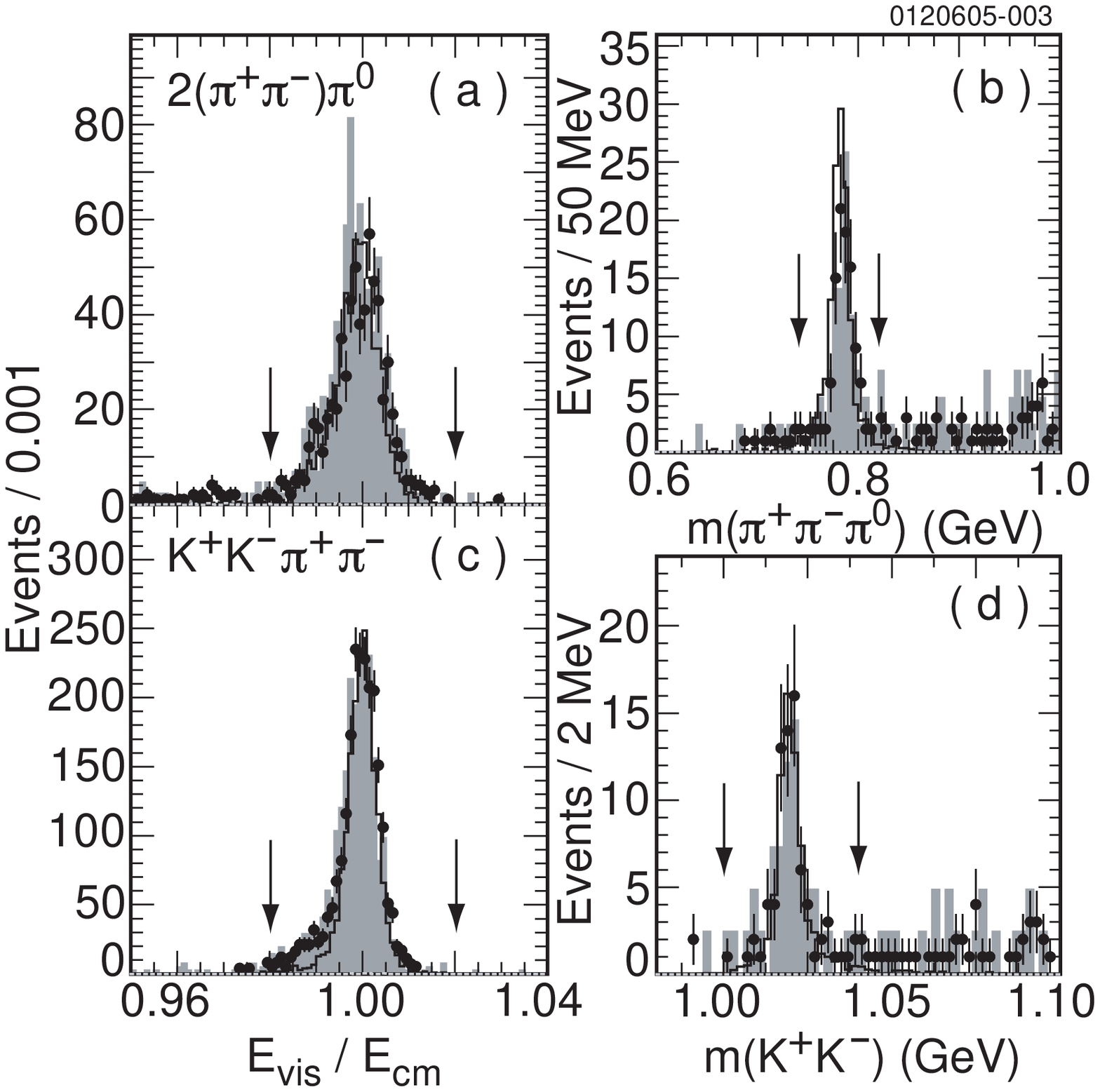}
  \caption{Distributions for the modes \FivePi\ [(a) and (b)],  
           \KKPiPi\ [(c) and (d)].
           The pairs of arrows indicate the signal regions.
           (a) and (c) The scaled total energy.
           (b) The $\pi^+ \pi^- \pi^0$ invariant mass in \FivePi.
           (d) The $K^+K^-$ invariant mass in \KKPiPi.
           Filled circle with error bar: $\psi(3770)$ data,
           solid line: $\pdp\to$ \FivePi\ or \KKPiPi\ Monte Carlo,
           dashed line: $\pdp\to\DDbar\to$ \FivePi\ or \KKPiPi\ Monte Carlo,
           shaded histogram: $e^+e^- \to \gamma^* \to$ \FivePi\ or \KKPiPi\ 
           from continuum data. }
  \label{fig:selected}
\end{figure}

In Fig.~\ref{fig:selected}, the scaled energy and invariant mass
distributions are shown for two typical modes: \FivePi\ and \KKPiPi. 
Evidence for production of the $\omega$ and $\phi$ resonances,
respectively, is observed in the corresponding mass spectra.
The background from $\DDbar$ is tiny, and thus is almost invisible.

Event totals are shown for both the $\pdp$ and the continuum 
in Table \ref{tab:num}, where $S_{\pdp}$ ($S_{\rm co}$) is the number 
of events in the signal region and  $B_{\pdp}$ ($B_{\rm co}$) the number 
of events in the sideband region in $\psi(3770)$ (continuum) data.
Also shown are yields for a $\DDbar$ Monte Carlo sample corresponding
to 10 times the integrated luminosity of the data:
$S_{\DDbar}$ ($B_{\DDbar}$) in the signal (sideband) region.
Under the assumption that interference between $\psi(3770)$ decay
and continuum production of the same final state is absent, 
the number of events attributable to each $\psi(3770)$ decay mode,
$N_S$, is
\begin{widetext}
\begin{equation}
N_S = S_{\pdp} - B_{\pdp} - f_{\rm co} (S_{\rm co}-B_{\rm co})
       - f_{\DDbar} (S_{\DDbar}-B_{\DDbar}),
\label{equ:ns}
\end{equation}
\end{widetext}
where $f_{\rm co}$ is mode dependent and listed in Table \ref{tab:num}, 
and $f_{\DDbar}$=0.1 is the scale factor for $\DDbar$ MC.
Since no statistically significant excess is observed, 
we obtain a 90\% C.L. upper limit on the number of events by
adding 1.64 times the statistical uncertainty determined 
from the yields on the continuum and at the $\pdp$.

The efficiency, $\epsilon$, for each final state is obtained 
using a MC simulation~\cite{GEANT} of the CLEO-c detector
based on the EvtGen event generator \cite{EvtGen}.
No initial state radiation is included in the Monte Carlo, 
but final state radiation is accounted for. The efficiencies 
in Table \ref{tab:num} include the branching ratios for 
intermediate final states.

\begin{table*}[!h]
\caption{For each final state $h$ the following quantities are given: 
the number of events in the signal region, $S_{\psi(3770)}$, 
and background from sidebands, $B_{\psi(3770)}$, in $\psi(3770)$ data; 
the number of events in the signal region, $S_{\rm co}$, 
and background from sidebands, $B_{\rm co}$, in continuum data; 
the scale  factor, $f_{\rm co}$; the number of events in 
the signal region, $S_{\DDbar}$, and background from sidebands, 
$B_{\DDbar}$, in a $\DDbar$ MC sample corresponding to 10 times the 
integrated luminosity of the $\psi(3770)$ data sample; the number of events
attributable to $\psi(3770)$ decay, $N_S$, computed according to
Eq.~\ref{equ:ns}; the significance, in units of standard deviations; 
the efficiency, $\epsilon$; 
the cross section upper limit including the systematic error (90\% C.L.), and 
the branching ratio upper limit including the systematic error (90\% C.L.).
For $\eta 3 \pi$, the two decays modes $\eta \to \gamma\gamma$ $^a$
and  $\eta \to 3 \pi$ $^b$ are combined on line $\eta 3\pi$.
(The sign of the significance indicates an excess/deficit of events).}
\begin{center}
\begin{tabular}{|c|c|c|c|c|c|c|c|c|c|c|c|c|}  \hline
 mode & \multicolumn{2}{|c|}{continuum} & $f_{\rm co}$ 
      & \multicolumn{2}{|c|}{$10\times\DDbar$ MC} 
      & \multicolumn{2}{|c|}{$\psi(3770)$} & $N_S$
      & Sig. & $\varepsilon$ & $\sigma$ U.L. & $\cal B$ U.L. \\ \hline
 $h$  & $S_{\rm co}$ & $B_{\rm co}$ &  & $S_{\DDbar}$ & $B_{\DDbar}$ 
      & $S_{\psi(3770)}$ & $B_{\psi(3770)}$   &   & (\#$\sigma$) &
      & (pb) & ($10^{-4}$) \\ \hline
\FourPi   & 1471 & 28 & 2.49 &  1 & 13 & 3411 &  90 & -266.5 & -2.5 & 0.4305 &  8.7 & 11.2 \\
\FivePi   &  350 & 18 & 2.26 & 15 & 14 &  647 &  18 & -120.5 & -2.6 & 0.1990 &  8.2 & 10.6 \\
\EtaPiPi  &   15 &  0 & 2.57 &  0 &  0 &   41 &   1 &    1.5 &  0.1 & 0.0450 &  9.7 & 12.4 \\
\OmegaPiPi&   43 &  9 & 2.35 &  0 &  0 &  107 &  18 &    9.1 &  0.5 & 0.1638 &  4.6 &  6.0 \\
\EtaTriPiA&   27 &  2 & 2.61 &  8 &  0 &   67 &  11 &  -10.1 & -0.6 & 0.0688 &  4.5 &  5.8 \\
\EtaTriPiB&   20 &  9 & 2.64 &  2 &  1 &   62 &  23 &    9.8 &  0.6 & 0.0248 & 24.0 & 30.7 \\
\EtaTriPi &      &    &      &    &    &      &     &        &      &        & 10.9 & 13.4 \\
\EtapTriPi&    1 &  0 & 2.75 &  1 &  0 &    5 &   0 &    2.2 &  0.4 & 0.0149 & 19.2 & 24.4 \\
\KKPiPi   &  954 & 25 & 2.40 & 32 &  7 & 2262 &  47 &  -16.8 & -0.2 & 0.3720 &  7.0 &  9.0 \\
\PhiPiPi  &   33 & 13 & 2.43 &  0 &  0 &   77 &  25 &    3.3 &  0.2 & 0.1629 &  3.2 &  4.1 \\
\PhiFzero &   12 &  5 & 2.49 &  0 &  2 &   32 &  15 &   -0.2 &  0.0 & 0.0863 &  3.5 &  4.5 \\
\KKTriPi  &  634 & 18 & 1.73 & 30 & 21 & 1121 &  32 &   24.9 &  0.5 & 0.1283 & 18.4 & 23.6 \\
\EtaKK    &    3 &  0 & 2.50 &  0 &  0 &    3 &   0 &   -4.5 & -0.7 & 0.0389 &  3.2 &  4.1 \\
\OmegaKK  &   62 & 12 & 2.31 &  0 &  1 &  114 &  14 &  -15.3 & -0.7 & 0.1269 &  2.6 &  3.4 \\
\KKKK     &  100 & 11 & 2.67 &  9 &  1 &  267 &   7 &   21.7 &  0.7 & 0.3170 &  4.6 &  6.0 \\
\PhiKK    &   46 & 15 & 2.59 &  4 &  0 &  118 &  22 &   15.2 &  0.7 & 0.1564 &  5.9 &  7.5 \\
\KKKKPi   &   20 &  0 & 2.88 &  8 &  0 &   50 &   0 &   -8.4 & -0.6 & 0.1479 &  2.2 &  2.9 \\
\PPPiPi   &  337 & 28 & 2.47 &  0 &  0 &  851 &  60 &   28.6 &  0.5 & 0.5149 &  4.5 &  5.8 \\
\PPTriPi  &  204 &  9 & 2.58 &  0 &  0 &  604 &  16 &   85.4 &  2.1 & 0.2259 & 14.4 & 18.5 \\
\EtaPP    &    2 &  1 & 2.62 &  0 &  0 &    4 &   2 &   -0.6 & -0.1 & 0.0469 &  4.2 &  5.4 \\
\OmegaPP  &   26 &  4 & 2.58 &  0 &  0 &   54 &   5 &   -7.8 & -0.5 & 0.1421 &  2.2 &  2.9 \\
\PPKK     &   25 &  1 & 2.62 &  0 &  0 &   89 &   3 &   23.0 &  1.5 & 0.4111 &  2.5 &  3.2 \\
\PhiPP    &    2 &  3 & 2.69 &  0 &  0 &    2 &   2 &    0.0 &  0.0 & 0.1872 &  1.1 &  1.3 \\
\LamLam   &    4 &  1 & 2.69 &  0 &  0 &    6 &   0 &   -2.1 & -0.3 & 0.2154 &  1.0 &  1.2 \\
\LLPiPi   &   23 &  4 & 2.37 &  0 &  0 &   42 &   7 &  -10.0 & -0.7 & 0.1019 &  2.0 &  2.5 \\
\LamPK    &   65 &  7 & 2.57 &  0 &  0 &  150 &  11 &  -10.0 & -0.4 & 0.2602 &  2.2 &  2.8 \\
\LamPKPiPi&   29 &  3 & 2.64 &  0 &  0 &   94 &  17 &    8.2 &  0.4 & 0.1471 &  4.9 &  6.3 \\
\hline
\end{tabular}
\label{tab:num}
\end{center}
\end{table*}

We correct the number of events by the efficiency $\varepsilon$, 
and normalize to the integrated luminosity $\cal L$ to obtain 
the cross section:
$$\sigma=\frac{N_S} {\varepsilon {\cal L}},$$
and normalize to the total number of the $\pdp$ decays to obtain
the branching ratio:
$${\cal B}=\frac{N_S} {\varepsilon N_{\pdp}}.$$
The number of the $\pdp$ decays is determined by
$N_{\pdp} = \sigma_{\rm tot} \cdot {\cal L},$
where $\sigma_{\rm tot}=(7.9\pm0.6)$ nb is the world average total cross 
section of the $\pdp$ from \cite{rosner}.

The systematic uncertainties on the ratio of branching fractions share
common contributions from the integrated luminosity (1.0\%),
uncertainty in $f_{\rm co}$ (10.0\%), trigger efficiency (1.0\%), and Monte
Carlo statistics (2.0\%). Other sources vary by channel. We include
the following contributions for detector performance modeling
quality: charged particle tracking (0.7\% per track), $\pi^0/\eta
(\to\gamma\gamma)$ finding (4.4\%), $\Lambda$ finding (3.0\%), $\pi / K/ p$
identification $(0.3\%/1.3\%/1.3\% {\rm ~per~identified~} \pi / K
/p)$, and scaled energy and mass resolutions (2.0\%). 
The systematic uncertainty associated with the sideband background
is obtained by coherently increasing the backgrounds both at the
$\pdp$ and on the continuum by one statistical sigma.
Since the background in many modes is small, the
Poisson probability for the observed number of background events
to fluctuate up to the 68\% C.L. value is calculated and
interpreted as the uncertainty in the level of background. 
Many of the modes studied have resonant submodes, however
the efficiencies do not differ by much. We generate MC data 
with a phase space model, and take a 10.0\% uncertainty for 
decay model dependence.
In the computation of the branching fractions, a common uncertainty
of 7.6\% enters due to the number of $\pdp$ decays arising from 
the uncertainty in the total cross section of the $\pdp$ \cite{rosner}.
We give the significance for each mode and upper limits 
(including the systematic error) for the cross section and 
branching ratio in Table \ref{tab:num}. 

In summary we have searched for 25 exclusive multibody hadronic 
decay modes at the $\pdp$.
No significant signal is observed  in any mode.
For each mode we give the significance, and the upper limit
on the cross section and branching fraction at 90\% C.L. 
This study together with the $\pp$ multibody decay analysis \cite{psipmulti}
provide useful information about $S-D$ mixing.
The cross section deficit remains a puzzle. 
However the uncertainty in the total cross section of the $\pdp$ is large. 
A fine energy scan over the $\pdp$ resonance to measure the total
cross section would be very valuable.

We gratefully acknowledge the effort of the CESR staff
in providing us with excellent luminosity and running conditions.
This work was supported by the National Science Foundation
and the U.S. Department of Energy.


\begin{thebibliography}{99}

\bibitem{PDG} S.~Eidelman {\it et al.} (Particle Data Group), 
{\Journal\PLB&592&1(2004)}.

\bibitem{rosner} J. Rosner, arXiv:hep-ph/0405196.

\bibitem{ddxsec} Q. He {\it et al.} (CLEO Collaboration), 
{\Journal\PRL&95&121801(2005)}.

\bibitem{bespipijpsi} J.~Z.~Bai {\it et al.} (BES Collaboration),
{\Journal\PLB&605&63(2005)}.

\bibitem{bkhhmk} N. E. Adam {\it et al.} (CLEO Collaboration),
arXiv:hep-ex/0508023.

\bibitem{YELLOWBOOK} R. A. Briere {\it et al.}
(CLEO-c/CESR-c Taskforces \& CLEO-c Collaboration), 
Cornell University LEPP Report No. CLNS~01/1742 (2001) (unpublished).

\bibitem{cleoiiidetector} Y. Kubota {\it et al.} (CLEO Collaboration), 
{\Journal\NIMA&320&66(1992)};
D. Peterson {\it et al.}, {\Journal\NIMA&478&142(2002)};
M.~Artuso {\it et al.}, {\Journal\NIMA&554&147(2005)}; 

\bibitem{LUMINS} G.~Crawford {\it et al.} (CLEO Collaboration),
{\Journal\NIMA&345&429(1994)}.

\bibitem{BBY} C.~M.~Carloni Calame {\it et al.},
  Nucl. Phys. B, Proc. Suppl. {\bf 131}, 48 (2004).

\bibitem{GEANT} Computer code GEANT~3.21, in R.~Brun {\it et al.}, CERN
Report No. W5013, (1993) (unpublished).

\bibitem{psipmulti} R. A. Briere {\it et al.} (CLEO Collaboration), 
{\Journal\PRL&95&062001(2005)}.

\bibitem{kinefit} As charged particle momentum resolution is excellent, 
a kinematically constrained fit only significantly improves the $\pdp$
mass resolution for final state particles that decay to all neutral 
final states.

\bibitem{EvtGen} D.J. Lange, {\Journal\NIMA&462&152(2001)}.

\end{thebibliography}
\end{document}